\documentclass[twocolumn,showpacs,preprintnumbers,amsmath,amssymb]{revtex4}
\usepackage{graphicx}
\usepackage[dvips]{epsfig}
\usepackage{xcolor}

\begin{document}

\title{Aging of colloidal gels in microgravity}
\author{Swagata S. Datta}
\affiliation{Beninngton College, Bennington, VT 05201, USA}
\author{Waad Paliwal}
\author{Eric R. Weeks}
\affiliation{Department of Physics, Emory University,
Atlanta, GA 30322, USA}
\email{erweeks@emory.edu}

\date{\today}

\begin{abstract}
We study the aging of colloidal gels using light microscopy movies of depletion gels from the International Space Station. Under such microgravity conditions, we observe a slowdown in particle dynamics consistent with gel aging.  Stronger attractive forces promote the formation of thicker gel strands over time.  The samples are bidisperse, composed of particles with a size ratio 1.2.  Larger particles experience stronger depletion forces, which lead to a large first-neighbor peak in the pair correlation function $g(r)$ due to the prevalence of large-large particle contacts. As the gel ages, small mobile particles are incorporated into the gel structure.  The changes in gel structure correlate with a slow power-law decay in particle motion, observed over nearly two orders of magnitude of time scales in microgravity experiments.
Additionally, through complementary ground-based experiments, we compare two-dimensional (2D) and three-dimensional (3D) images of depletion colloidal gels. While microgravity gel data are limited to 2D projections, ground-based data establish a correspondence between the 2D and 3D $g(r)$ peak heights.  Our results provide new insights into how colloidal gels age in the absence of gravitational collapse.

\end{abstract}

\maketitle



\section{Introduction}

Colloids are mixtures where small solid particles are suspended in a liquid medium. When attractive forces are present between these particles, they can adhere to each other, forming a colloidal gel \cite{anderson02,dawson02,poon02,Royall21}. In such gels, particles aggregate into free-floating aggregates \cite{segre01prl,sciortino04,ohtsuka08} or large tendrils that span the entire system; this latter state is termed a ``colloidal gel'' \cite{lu08,mohraz08,joshi14,rouwhorst20}.  This unique microstructure gives colloidal gels distinct properties, making them valuable in various everyday products, including food, paint, cosmetics, and pharmaceuticals \cite{Royall21}. Due to their wide range of applications, extensive research has been dedicated to understanding the structure, dynamics, and mechanical properties of colloidal gels. This knowledge proves to be essential to predict how well products can preserve their characteristic properties throughout their shelf life \cite{Dipjyoti10}. 

Gels are inherently out of equilibrium, and so their properties can change over time, making them useful model systems for studying nonequilibrium physics \cite{joshi14}.  An initial state of a colloidal system is individual particles freely diffusing, where perhaps the system is initialized by changing conditions so that the particles become attractive \cite{cipelletti00}.  The earliest time scale is the Brownian time scale characterizing the typical time for a particle to diffuse its own radius. After the initialization, over some time scale proportional to the Brownian time scale, particles come into contact due to their diffusive motion and may stick together. This early stage of gel formation is closely related to the phenomenon of spinodal decomposition \cite{verhaegh97,lietorsantos10,royall18,Royall21}.  The early stage of gel formation is also often described in terms of diffusion-limited cluster aggregation or reaction-limited cluster aggregation \cite{eggersdorfer_structure_2012,chou_effects_1996,anderson01}.

On longer time scales, the initial structure of the gel has formed -- defined by a system-spanning network of colloidal gel strands -- but nonetheless, the structure of the gel continues to slowly evolve.  The slow structural changes typically correlate with changes in the mechanical properties of the gel \cite{Royall21,manley05}.  The slow evolution of the gel is often termed ``aging'' or ``coarsening'' and takes place over exponentially growing time scales.  Generally, ``aging'' can refer to any system which is out of equilibrium with the ``age'' as the time $t$ elapsed since the initial preparation \cite{hunter12rpp,joshi14}.  In this paper we use ``aging'' to refer to any changes of a colloidal gel over time.  In contrast, we will use ``coarsening'' to refer specifically to structural changes such as the gel strands becoming thicker, or more particles being incorporated into the gel \cite{Calzolari17,zia14}.  Both coarsening and aging involve changes in structure and dynamics:  to change the structure, particles have to move; as the structure changes, the ability of particles to move is affected.  In general, the goal of studying these slow changes is to understand the behavior of colloidal gels over a several orders of magnitude of time scales, with special emphasis on understanding the $t \rightarrow \infty$ asymptotic behavior.


Prior work examining the temporal evolution of colloidal gels has included microscopy studies \cite{verhaegh97,verhaegh99,dehoog01,royall18} and light scattering studies \cite{poon95,poon97,verhaegh97,cipelletti00}.  Microscopy found a slight coarsening of colloidal clusters with increasing age \cite{verhaegh97}.  In one microscopy study, fissures opened up in the gel, although these may have been caused by gravitational effects \cite{verhaegh99}.  Confocal microscopy studies have identified local structures which change within colloidal gels as the gels coarsen:  for example, in some samples, the number of particles with exactly five neighbors increased with time \cite{royall18}.
Light scattering studies found a slowing of the dynamics as the gel ages, with motion due to gel restructuring; the causes of the slowing had to be inferred from the wave-vector dependence \cite{poon95,poon97,cipelletti00}.  Light scattering is also useful for quantifying the fractal dimension of colloidal gels \cite{piazza94,poon97}.  A rheology experiment also studied the coarsening of colloidal gels, finding that the elastic modulus increased with the age of the sample \cite{manley05}.

However, studies of the long-time aging of colloidal gels are complicated by gravity.  Gravity leads to sedimentation, changing the long-time properties of a colloidal gel \cite{buscall87,poon95,dehoog01}.  This arises because in general colloidal particles have not the same density as the background fluid in which they are suspended.  For a single particle, sedimentation is straightforward to quantify \cite{jones02,hunter12rpp}.  There is a gravitational force that depends on the density difference $\Delta \rho$ between the particle and the liquid, and this is balanced by the viscous drag force exerted on a moving particle.  The steady-state
sedimentation velocity can expressed as: 
\begin{equation}
\label{sediment}
v_{\rm sed} = \frac{2}{9} \frac{\Delta \rho g a^2}{\eta}.
\end{equation} 
in terms of the gravitational acceleration $g$, the density difference between the particle and the liquid medium $\Delta \rho$, and particle radius $a$ \cite{hunter12rpp}.  By forming a network of strands that extend to the walls of the container, colloidal gels help to hold suspended particles against gravity \cite{padmanabhan18}. However, Eq.~\ref{sediment} illustrates the challenge: a cluster of particles has a greater effective radius $a$ and thus is more strongly pulled by gravity. Given that the sedimentation velocity $v_{\rm sed}$ scales as $a^2$, a larger chunk of colloidal gel will fall even faster than isolated particles. This results in the gravitational collapse of colloidal gels \cite{starrs99,lietorsantos10,padmanabhan18}.  A significant number of studies focused on understanding the details of gravitational collapse, since this is the natural behavior in a laboratory \cite{buscall87,poon95,starrs99,verhaegh97,verhaegh99,dehoog01,starrs02,manley05collapse,huh07,bartlett12,teece14,Razali17,Filiberti19}. There are two distinct types of gravitational collapse: rapid collapse that occurs as the gel is prepared \cite{Razali17} and delayed collapse that takes place after a considerable time \cite{starrs99,starrs02,huh07,buscall09,bartlett12,teece14}. The latter typically leads to a catastrophic failure of the gel and a drastic change of the material's mechanical properties; this can limit the shelf-life of colloidal gel-based materials.

Given that gravitational collapse limits the ability of experimentalists to study gel coarsening and aging at long time scales, steps can be taken to mitigate the problem.  One approach motivated by Eq.~\ref{sediment} is to decrease $\Delta \rho$, that is, to match the density of the solvent to the density of the solid colloidal particles.  While widely used, this method is limited by the fact that perfect density matching is impossible to achieve. The challenge of density matching increases if one wants to do optical microscopy studies: the particle radius $a$ needs to be at least $\sim 0.2$~$\mu$m to enable particle-scale study of structure and dynamics within a colloidal gel.  Given that $v_{\rm sed}$ scales as $\Delta \rho a^2$, larger particles make it harder to achieve effective density matching.

\begin{figure}[t]
\centerline{
\epsfxsize=8.0truecm\epsffile{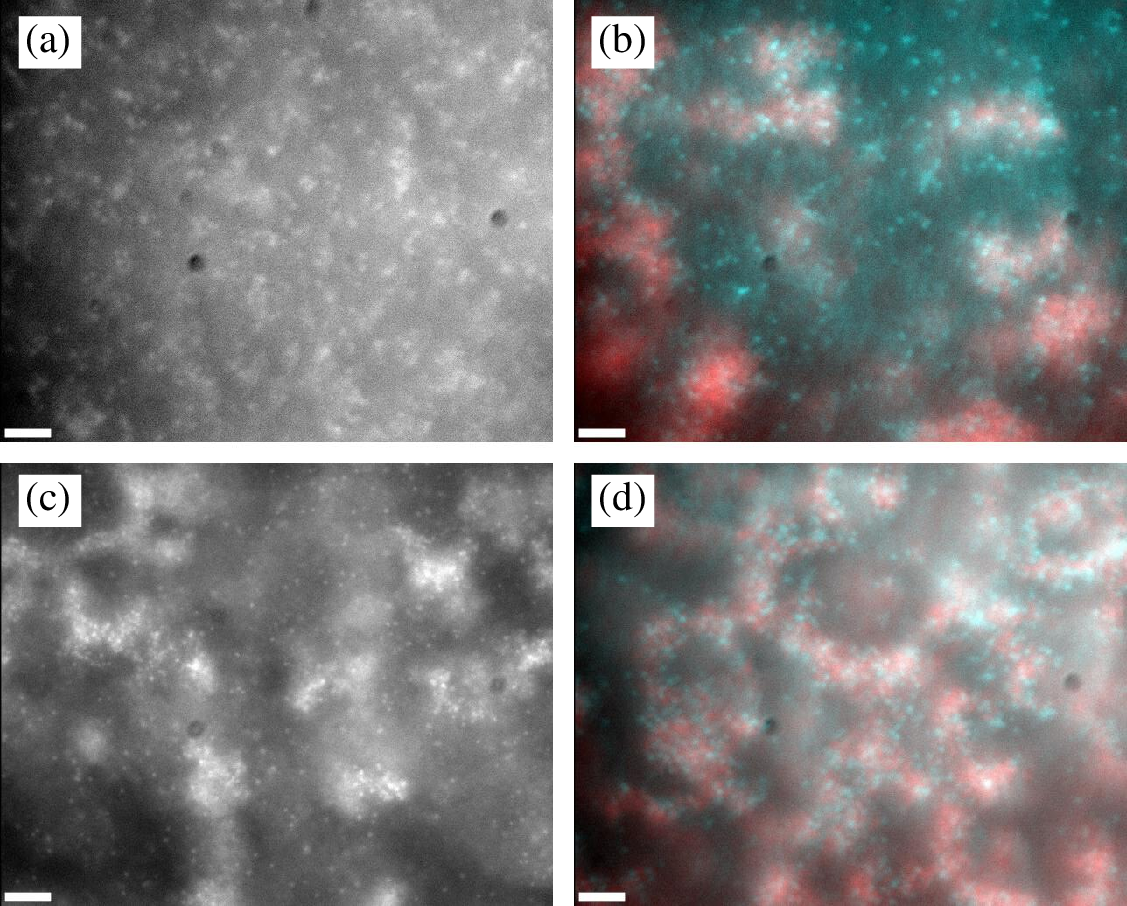}}
\caption{\small Images of a colloidal gel from ACE-M-1 
obtained from the PSI database.  The contrast is adjusted in these
images by the authors.  Scale bars indicate 10~$\mu$m.  (a) This image shows the larger particles (radius $a_{\rm L} = 1.1$~$\mu$m), lowest polymer concentration $c=0.5$~mg/ml, at time $t=2$~hr.  
(b) A gel with $c=0.9$~mg/ml at $t=12$~hr.  Image rendered in color by the authors.  The red component is the larger particles, and the teal is the smaller particles. 
(c) This image shows the smaller particles (radius $a_{\rm S} = 0.9$~$\mu$m), $c=1.2$~mg/ml, at $t=56$~hr.  (d) A gel with $c=2.0$~mg/ml at $t=2$~hr, rendered in color.  There are three black circles visible in all images, most especially in (a), which are likely specks of dirt on the microscope optics as they are stationary in time and appear identical in all images.
}
\label{gelpic}
\end{figure}

A second approach to mitigate the effects of sedimentation is to use microgravity: to reduce $g$ in Eq.~\ref{sediment}.  This is precisely the aim of the NASA ACE-M-1 (Advanced Colloids Experiment), conducted in 2013-2014. Proposed by Matthew Lynch (Proctor and Gamble) and Thomas Kodger (Harvard University/University of Amsterdam), ACE-M-1 used the Light Microscopy Module on the International Space Station to take microscopy videos of colloidal gels under microgravity conditions; see Fig.~\ref{gelpic} for example images. This experiment resulted in hundreds of thousands of images.

In this paper we use the ACE-M-1 microscopy videos to study the temporal evolution of several different colloidal gels under microgravity conditions.  Our chief observations are that these colloidal gels coarsen as they age:  their properties slowly change with time, with the observation time reaching more than 50 hours.  The temporal changes include both structural changes (more particles stick together as time passes) and dynamical changes (particle Brownian motion becomes more constricted as particles bind into the gels).  As the ACE-M-1 images are two-dimensional (2D) cuts through three-dimensional (3D) samples, we additionally take new confocal microscope images of colloidal gels to quantify the differences between 2D and 3D images.  While structural properties from 2D images differ from those same properties measured from the truer 3D images, we demonstrate that the observations between 2D and 3D are correlated.  Overall, the microgravity observations allow us to study the particle-scale details of these aging colloidal gels over time scales ranging from seconds to tens of hours.

As noted above, on Earth sedimentation causes colloidal gels to collapse, making long-duration studies such as ours challenging or impossible.  That being said, Buscall {\it et al.} have pointed out that for delayed onset gravitational collapse, before the collapse the gel is aging and coarsening, changing the gel so that at some point the structure suddenly collapses \cite{buscall09}.  Thus, our results are likely relevant for Earth-based colloidal gels on time scales intermediate between the initial gel formation and the subsequent gravitational collapse.

\section{Experimental methods and data}

\subsection{NASA microgravity data}

Our project primarily involves analyzing data from NASA’s Physical Systems Informatics (PSI) database \cite{nasapsiwebsite}.  In particular, we are analyzing microscope images of colloidal gels from the ACE-M-1 experiment (Advanced Colloids Experiment).  ACE-M-1 studied a series of colloidal samples that could form gels.  The particles were 27\% sulfolane (poly-TFEMA) and 73\% formamide (poly-TBMA), and they were dispersed in a solvent that was a mixture of 25 wt\% tetramethylsulfone, 75 wt\% formamide.  To cause the particles to stick together, 1 MDa polystyrenesulfonate was added as a polymer depletant, as is commonly done to form colloidal gels \cite{verhaegh97,starrs99,dehoog01,anderson02,poon02,dinsmore06,laurati09}.

The strength of the interaction potential increases with polymer concentration. The ACE-M-1 experiments studied ten samples with the polymer concentration varying by a factor of four from the smallest to largest.  In particular, the lowest polymer concentration was intended to induce only minimal attractive forces (no colloidal gelation), and the highest polymer concentration was intended to form a very sticky colloidal gel (all particles captured into the gel).  Indeed, this is what was observed.

One additional detail is that the colloidal samples were a mixture of small and large particles:  $a_{\rm S}=0.9$~$\mu$m and $a_{\rm L}=1.1$~$\mu$m in radii, equal numbers of each type of particles.  Each particle species has a polydispersity of $\sim 9$\% (standard deviation of the radius divided by the mean radius).  The depletion force is greater for the larger particles by a factor of $a_{\rm L}/a_{\rm S} = 1.49$ \cite{asakura54,kaplan94,dinsmore96,yodh01}.  The two particles were also dyed with different fluorescent dyes, such that the fluorescence microscope can distinguish the two sizes, see Fig.~\ref{gelpic}(b,d).   White portions of these images indicate crosstalk between the two color channels; the microscope fluorescence filter cubes do not completely separate the particle types.

The ACE-M-1 experiment resulted in 530 GB of images.  Each image is $1264 \times 1012$ pixels with $0.095$~$\mu$m/pixel.  The movies of ACE-M-1 were taken at 1 frame per second using fluorescence microscopy.  Each hour the experiment acquired 500 images of the small particles followed by 500 images of the large particles.  This cycle was repeated every hour for several days (typically, 50-60 total hours).  Some of the movies have flaws:  unusable images (with video camera distortions) or missing images; thus in some of the results presented, there are not data points for every hour.  For the results of this work, we analyze approximately 240,000 microscope images.  We use standard image analysis techniques to identify the particle positions in each image and then track the particles over time \cite{crocker96}.  Due to the use of fluorescence microscopy rather than confocal microscopy, the image quality is not ideal, and thus we can identify particle positions in each image to an accuracy of $0.1 - 0.2$~$\mu$m depending on image quality.  Tracking particles depends on the particles having displacements between video frames that are typically smaller than the interparticle separation distance.  We find this condition is met for all samples.  While the original goal of the ACE-M-1 experiment was to study ten different samples, as just discussed the data sets are of varying quality, and we find four data sets of suitable quality for analysis.  They are listed in Table I.  As noted above, the different samples are made with different levels of attractive interaction.  The sample with the lowest polymer concentration of $c=0.5$~mg/ml and thus the least attractive forces did not form a gel, and serves as a useful control experiment.

\begin{table}[tbh]
\begin{center}
\begin{tabular}{ c| c| c}
 Sample number & $c$ (mg/ml) & NASA identification\\ 
 \hline
 1 & 2.0 & GMT203 \\  
 5 & 1.2 & GMT274 \\
 8 & 0.9 & GMT175 \\
 10 & 0.5 & GMT219 \\
 \hline
\end{tabular}
\end{center}
\caption{Listing of the four NASA experiments analyzed in this work.  The first column is the sample number according to the original experiment.  The second column gives the polymer concentration $c$.  The ``NASA identification'' entry corresponds to how the sample is tagged in the NASA PSI database.}
\label{sampletable}
\end{table}

\subsection{Ground-based experiments} 
\label{sec:groundmethods}

To complement the analysis of the ACE-M-1 results, we conducted analogous ground-based experiments. Similar to the ACE-M-1 gels, we produced colloidal gels with a polymer depletant. We use colloidal poly-methyl-methacrylate (PMMA) particles obtained from Andrew Schofield and Wilson C. K. Poon from the University of Edinburgh. The PMMA particles are sterically stabilized with a hard core and are slightly charged. The particle radius is 1.08~$\mu$m with a polydispersity of 5\%. We use polystyrene for the depletant (molecular weight $M_{w} = 3,085,000$~g/mol, Polymer Laboratories Inc.~UK.). The goal of the ground experiments is to obtain a variety of gel samples. We produce 13 distinct samples with polymer concentrations spanning from 6.9~mg/ml to 16~mg/ml and particle volume fraction $\phi = 0.03-0.29$. The particles are suspended with a mixture of cyclohexyl bromide (CXB) and decalin (DCL) at a 85/15 (w/w) ratio. This particular mixture has two useful properties: it matches the particle density to minimize sedimentation, and it matches the particle index of refraction which is useful for the microscopy \cite{dinsmore01}.

\begin{figure}
\centering
\includegraphics[width=0.7\columnwidth]{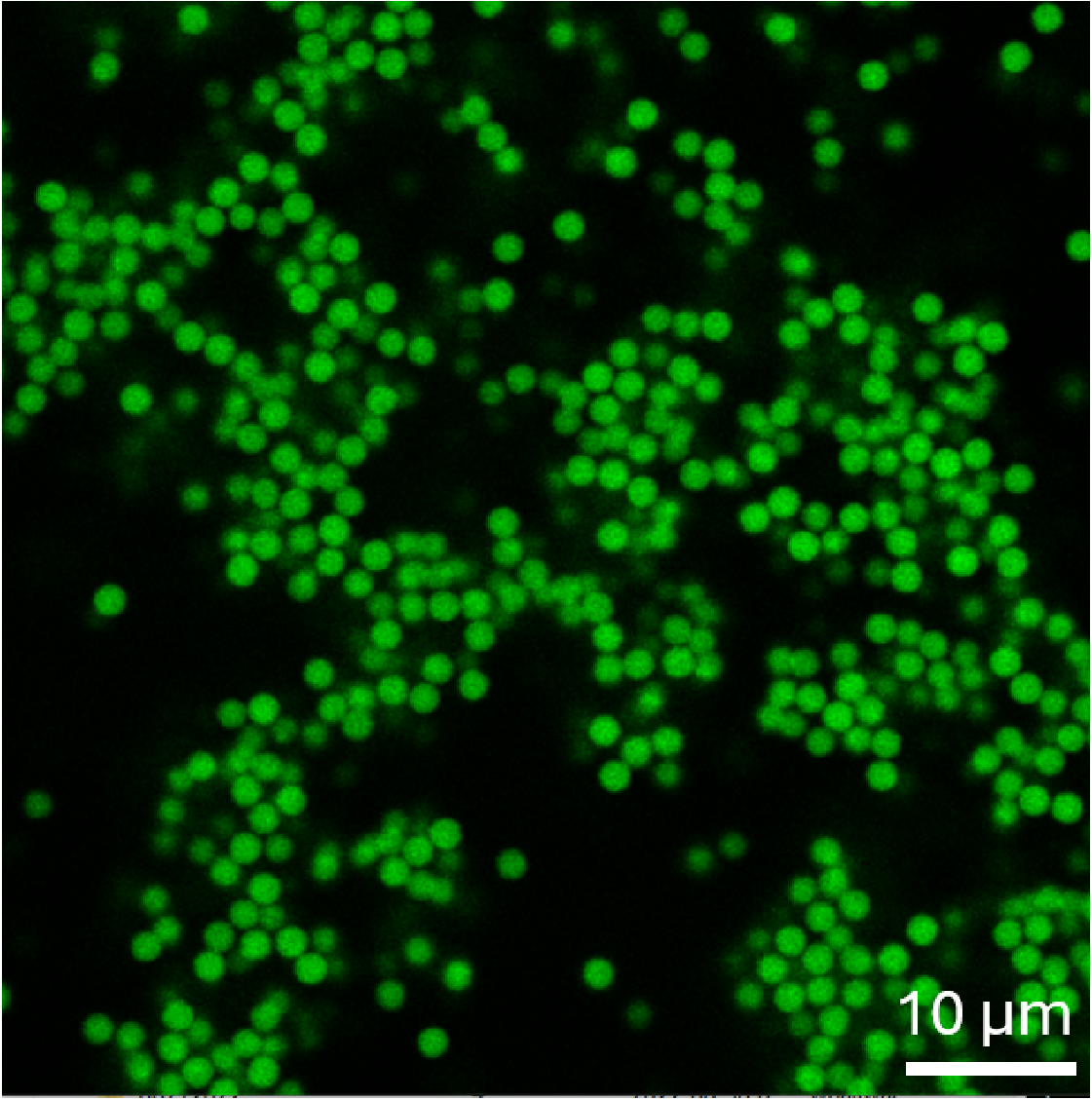}
\caption{\label{gel}
Confocal microscope image of a colloidal gel obtained from the ground experiments. The volume fraction is $\phi = 0.11$ and the depletant concentration is $c=8.5$~mg/ml.
}
\end{figure}

\begin{figure}
\centering
\includegraphics[width=0.8\columnwidth]{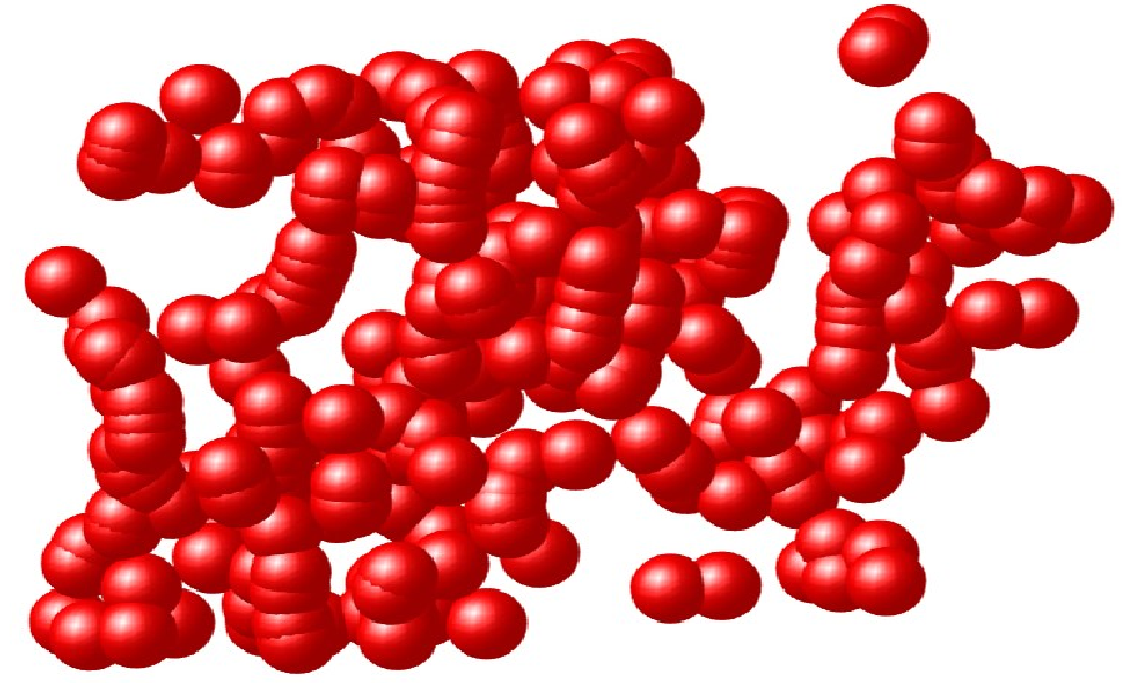}
\caption{\label{gel2}
A 3D rendering of a region of the gel from Figure \ref{gel} produced from the output of the particle tracking algorithm. The particle radius is 1.08~$\mu$m. 
}
\end{figure}

To image the gels, we use a Leica SP8 inverted fluorescence confocal microscope to produce 3D image stacks for each sample.  See Fig.~\ref{gel} for a 2D image from one such stack and Fig.~\ref{gel2} for a 3D rendering from the same sample.  The 3D images are $512 \times 64 \times 85$ voxels, with each voxel having dimensions $0.137 \times 0.137 \times 0.3$~$\mu$m$^3$.  We use a $63\times$ oil objective with a Numerical Aperture of 1.4.  The goal of these experiments is to compare the analysis between 2D and 3D images of the same sample.  In this case, the 2D images are simply the individual image slices comprising the 3D image stack.  The resultant 2D and 3D images are analyzed with standard particle tracking codes \cite{crocker96,dinsmore01}.  Here our particle position uncertainty is 0.1~$\mu$m in $x$ and $y$, and 0.3~$\mu$m in $z$.  By comparing our ground-based experimental 2D and 3D images, we aim to gain insight about the 3D structure of the ACE-M-1 gels based on their 2D images.  For all ground-based experimental results the volume fraction we report is as-measured in the 3D images using direct particle counting. This method is subject to systematic uncertainty related to the uncertainty of the mean particle radius which is $\pm 0.01$~$\mu$m \cite{royall13}.

\section{Aging results}

\subsection{The dynamics slow}

The ACE-M-1 samples are initialized by vigorous stirring, as is traditional for many colloidal gel studies.  The stirring was accomplished with a small teflon-coated stir bar inside the chamber.  The stirring breaks apart any preexisting gel and sets an initial time.  However, this initial time is rather uncertain: in some cases it was tens of seconds, and in others, as long as tens of minutes \cite{kodgerprivate}.  Fortunately, the data taken span hours:  so the uncertainty of $t=0$ is a small correction for aging times $t>1$~hr.  In the following, when we plot the data as a function of $t$, $t=1$~hr is defined as the first movie in a data set.

After stirring ceases, the colloidal particles then diffuse, and when they contact each other, they typically stick because of the polymer depletant.  Over time, the clusters become larger and then form a system-spanning network.  As the gel continues to age, remaining free particles can fill in and thicken the gel strands.  In addition, particles can occasionally detach and reattach elsewhere; the more contacting neighbors a particle has, the more stable its position is due to the higher contact energy \cite{puertas04,zia14}.  The slow evolution of the gel's structure and dynamics is termed aging, and the structural changes are specifically termed coarsening.

\begin{figure}
    \centering
    \includegraphics[width=0.9\columnwidth]{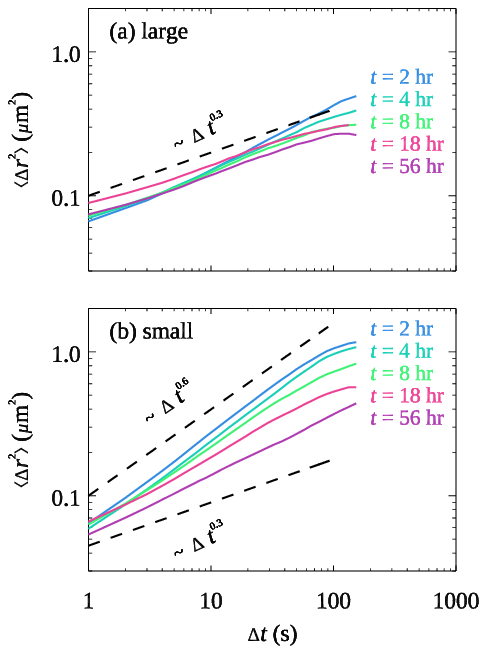}
    \caption{Mean square displacement of the (a) larger and (b) smaller particles within the sample with the highest depletion force (depletant concentration $c=2.0$~mg/ml).  The curves are calculated at the ages as indicated.  As the gel ages, the data shift downwards showing the slowing down of the movement. The dashed lines have power law slopes as labeled. }
    \label{fig:sample1msd}
\end{figure}

Aging is clearly apparent in the ACE-M-1 colloidal gel samples: the dynamics slow down as a function of time.  This can be seen in Fig.~\ref{fig:sample1msd}, which shows the mean square displacement (MSD) for a colloidal gel sample taken at different ages since the gel formation.  The curves rise with lag time $\Delta t$ showing that the particles diffuse slightly, although their motion grows more slowly with lag time than normal diffusion of colloidal particles, which would have a mean square displacement growing as $\Delta t^1$.  In the figure, especially at lag time $\Delta t \approx 100$~s, the dynamics are noticeably slower as the sample ages from 2 hours to 56 hours since preparation.  The slight increase in the mean square displacement at short lag times is related to photobleaching, which makes finding the particle centers slightly more uncertain, causing a fictitious rise in the short time MSD \cite{poon12}.

\begin{figure}
    \centering
    \includegraphics[width=0.9\columnwidth]{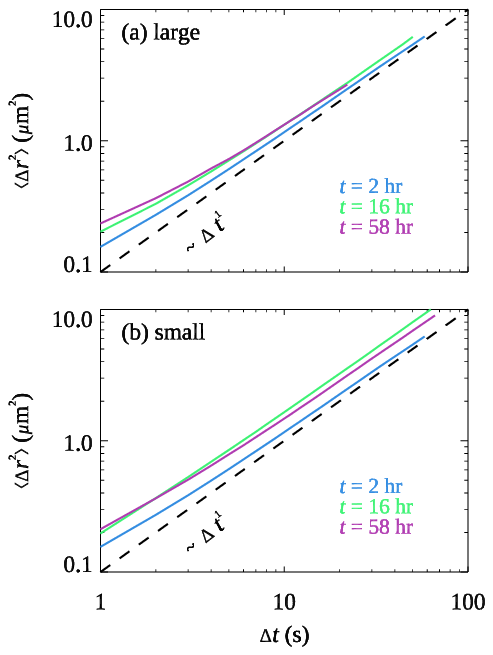}
    \caption{Mean square displacement of the (a) larger and (b) smaller particles within the sample with the least depletion force (depletant concentration $c=0.5$~mg/ml).  The curves are calculated at the ages as indicated. The dashed line indicates a power law slope of 1.0, consistent with normal diffusion.}
    \label{fig:sample10msd}
\end{figure}

The behavior of the colloidal gel shown in Fig.~\ref{fig:sample1msd} is in contrast to the sample with the least attractive interactions, for which the particles freely diffuse throughout the course of the experiment.  The mean square displacements for the low-attraction sample are shown in Fig.~\ref{fig:sample10msd}.  Here, the MSD curves do not change significantly with age $t$.  The fact that the MSD grows linearly with the delay time $\Delta t$ shows that these particles are freely diffusing for the duration of the experiment.

\begin{figure}
\centering
\includegraphics[width=0.95\columnwidth]{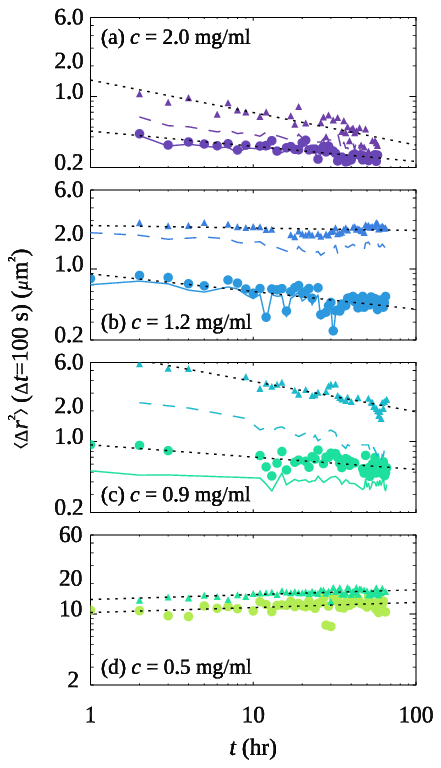}
\caption{\label{msd-aging}
Mean square displacement of the samples, evaluated with a lag time $\Delta t = 100$~s, as a function of the age of the gel.  The small triangles are for the small particles ($a_{\rm S}=0.9$~$\mu$m) and the large circles are for the large particles ($a_{\rm L}=1.1$~$\mu$m).  The colored dash lines and solid lines correspond to small and large particles, respectively, with the most mobile particles removed from the average.  The black dotted lines are fits to power-law decay, $\langle \Delta r^2 \rangle \sim t^\alpha$, with exponents $\alpha$ listed in Table II.  The panels correspond to different depletant concentrations as indicated.
}
\end{figure}

We quantify the changes in the samples using the MSD evaluated at lag time $\Delta t=100$~s.  This time scale is used as it is long enough that the particles make significant displacements typically greater than our particle positional uncertainty; but also it is shorter than the 500~s duration of the movies so that we can do some time-averaging within a given movie.  The results we obtain are not sensitive to this precise choice.  The results are shown in Fig.~\ref{msd-aging}.  For the three colloidal gel samples [panels (a) - (c)], the particle motion slows slightly with sample age.  The data decay consistent with a weak power law, $\langle \Delta r^2 \rangle \sim t^\alpha$, with decay exponents ranging from $-0.03$ to $-0.28$; specific values are given in Table~II.  In general, the stronger the depletion force, the less motion particles have \cite{dibble06}.  In contrast to the colloidal gels, the sample with the least depletion force barely shows any change with sample age, as shown in Fig.~\ref{msd-aging}(d).  The slight increase in the measured motion for this sample is likely due to an increase in particle positional uncertainty due to poorer image quality in these experimental movies at later times, which causes an increase in apparent motion \cite{poon12}.  The ratio of the data in (d), $\langle \Delta r^2_S \rangle / \langle \Delta r^2_L \rangle$, is equal to $a_L/a_S$ within the measurement error, as expected for particles undergoing normal diffusion as per the Stokes-Einstein-Sutherland equation for the diffusion constant ($D = k_B T / 6 \pi \eta a$ with Boltzmann constant $k_B$, absolute temperature $T$, solvent viscosity $\eta$, and particle radius $a$) \cite{sutherland1905,einstein1905a}.

\begin{figure}
\centering
\includegraphics[width=0.95\columnwidth]{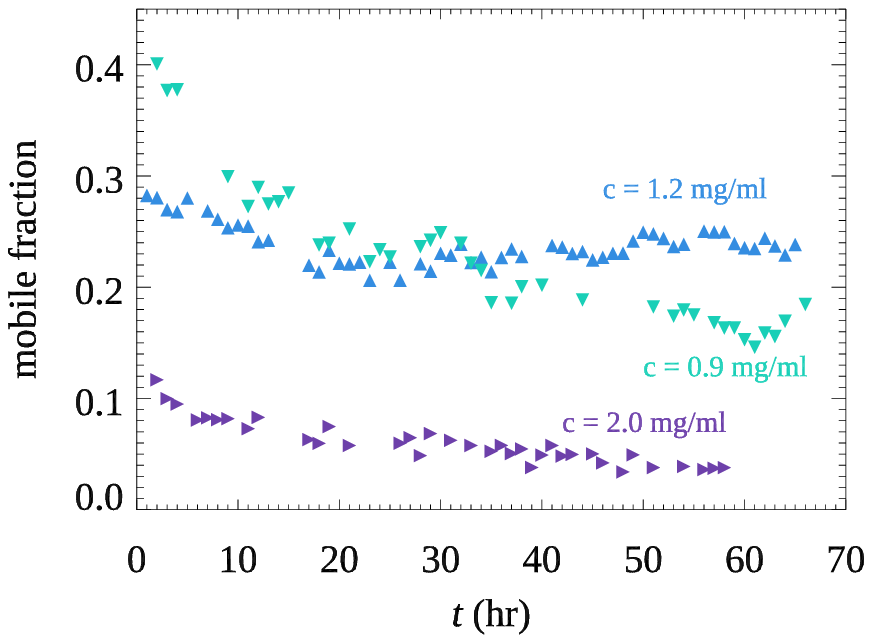}
\caption{\label{monomers}
Fraction of small particles that have displacements $\Delta r(\Delta t = 20$~s$) > d_{\rm S}/2$ as a function of age $t$, for the three gel samples as indicated.  For the non-gelling sample (with $c=0.5$~mg/ml), the mobile fraction is constant with age with a value of $0.75 \pm 0.01$ (mean $\pm$ standard deviation).
}
\end{figure}

Unattached and freely diffusing particles are visible in Fig.~\ref{gelpic}(b,c) (similar to some prior colloidal gel observations \cite{gao07}).  Some of the slowing down of the dynamics is due to these free particles attaching to the gel over time.  Most of the free particles are the smaller species.  To quantify this, we calculate the displacements of all small particles over a time scale $\Delta t = 20$~s.  This time interval is sufficient for the free particles to move more than the tracking uncertainty and, indeed, to move more than their radius $a_{\rm S}$; the bound particles move much less in this short time interval.  We then measure the fraction of small particles that have motion $\Delta r > a_{\rm S}$ over $\Delta t$.  The fraction of mobile particles as a function of time is plotted in Fig.~\ref{monomers}.  The sample with the lowest polymer concentration has the highest number of mobile particles initially, and vice versa for the sample with the highest polymer concentration.  The number of mobile particles decreases over time.  The sample with $c=0.9$~mg/ml crosses over the sample with $c=1.2$~mg/ml; both of those samples end with approximately 20\% of the particles being mobile.  Using the same criterion to define mobile large particles [$\Delta r(\Delta t = 20$~s$) > a_{\rm S}$], the same three samples initially have few large mobile particles (mobile fraction 0.1 for $c=0.9, 1.2$~mg/ml and 0.05 for $c=2.0$~mg/ml) and show slight decreases in this fraction over the duration of the experiment.  The larger particles experience a larger depletion force \cite{asakura54,kaplan94,dinsmore96}, so it is not surprising that they are bound to the gel structures more than the small particles.

While the decrease in number of freely mobile particles can explain some of the decrease in the dynamics, the other possibility is rearrangement within the gel strands \cite{zia14}.  To examine this question, we identify all mobile particles, including large particles that fit the definition ($\langle \Delta r (\Delta t = 20$~s$) > a_{\rm S}$), and remove them from the analysis of the dynamics over time.  We quantify the displacements of the remaining (bound) particles by again averaging all of their displacements over the original time interval $\Delta t=100$~s.  The resulting motion is shown in Fig.~\ref{msd-aging}(a,b,c) as the dashed lines (small bound particles) and solid lines (large bound particles).  The first observation to make is that the trend in motion is unchanged:  motion slows as the sample ages.  This shows that some of the slowing is not connected to the decreasing numbers of mobile particles.  The slope of the power-law decay of the motion is listed for all experiments in Table II, showing no systematic difference between analyzing all particles and analyzing just the bound particles.  

A second observation is that the gap between all the data (symbols) and the data of just bound particles (lines) is largest for the small particles; this matches the above observation that the smallest particles are much more likely to be mobile.  In contrast, the large particles are less likely to be mobile.  For the two highest polymer concentrations (and thus strongest attractive forces), there is barely any difference between the mobility of all large particles and bound large particles only [circles and solid lines respectively in Fig.~\ref{msd-aging}(a,b)].  For Fig.~\ref{msd-aging}(d), given that at $c=0.5$~mg/ml all particles are freely floating, the distinction between mobile and bound particles is moot.

\begin{table}[tbh]
\begin{center}
\begin{tabular}{ c| c c  c  c}
 $c$ (mg/ml) & $\alpha_{\rm L,all}$ & $\alpha_{\rm S,all}$ & $\alpha_{\rm L,immobile}$ & $\alpha_{\rm S,immobile}$ \\ 
 \hline
 2.0 & -0.15 & -0.32 & -0.15 & -0.23 \\  
 1.2 & -0.17 & -0.02 & -0.16 & -0.06 \\
 0.9 & -0.12 & -0.30 & -0.07 & -0.35 \\
 0.5 & +0.05 & +0.05 & n/a & n/a \\
 \hline
\end{tabular}
\end{center}
\caption{The power law decay exponents $\alpha$ for the data shown in Fig.~\ref{msd-aging}.  The subscripts indicate large (L) or small (S) particles, and whether all particles are used in the analysis or just the immobile particles.}
\label{powerlaw}
\end{table}

\subsection{Structure coarsens}

\begin{figure}
\centering
\includegraphics[width=0.8\columnwidth]{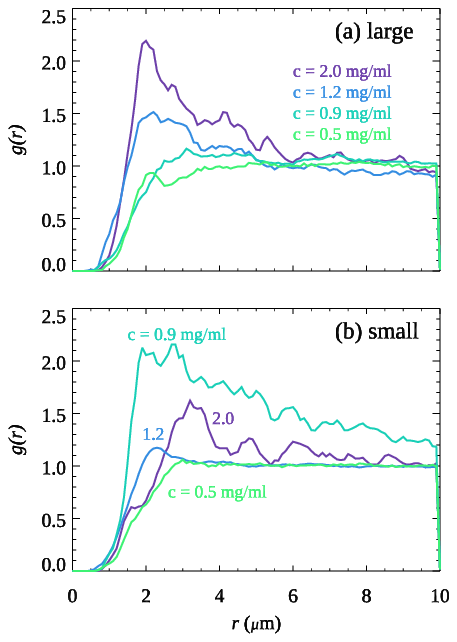}
\caption{\label{gofrnasa}
 Pair correlation function g(r) from four different samples during the first hour of observation, with the curves corresponding to samples with different depletant concentrations as indicated.  (a) Data for large particle correlations ($a_{\rm L}=1.1$~$\mu$m).  (b) Data for small particle correlations ($a_{\rm S}=0.9$~$\mu$m).  The prominent peak around $r = 2$~$\mu$m indicates that particles are sticking together; a larger peak means that more particles are stuck together.
}
\end{figure}

We wish to understand the structural changes which accompany the dynamical changes.  We start by measuring the pair correlation function $g(r)$ from the 2D images.  $g(r)$ relates to the likelihood of finding another particle at a distance $r$ from a given particle.  In this case, we examine the large-large particle correlations separately from the small-small particle correlations.  The results are shown in Fig.~\ref{gofrnasa} for (a) large and (b) small particles from the four samples we study, all taken during the first hour of observation.  The first peak of $g(r)$ is typically located at a distance corresponding to the particle diameter ($r \approx 2a$) and is related to the prevalence of particles in contact.  For large particles, the stronger depletion force (larger $c$) corresponds to a taller first peak, as expected.  For the small particles, this is not true; this is likely because many of the small particles are not stuck to each other, but rather are sticking to large particles.  As the small- and large-particle images are taken at different times, we cannot measure spatial correlations between the two particle species.

\begin{figure}
\centering
\includegraphics[width=0.8\columnwidth]{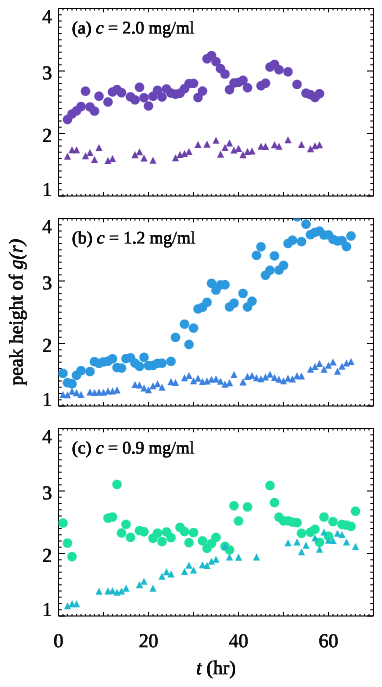}
\caption{\label{fig:grtime}
Peak height of the pair correlation function g(r) as a function of the age of the sample for three different samples with depletant concentrations $c$ as indicated.  The large circles correspond to the larger particles with radius $a_{\rm L}=1.1$~$\mu$m and the small triangles correspond to the smaller particles with radius $a_{\rm S}=0.9$~$\mu$m.  The fourth sample with $c=0.5$~mg/ml depleting concentration does not show any systematic changes with age (data not shown).
}
\end{figure}

Given that the peak heights of $g(r)$ reflect the degree of particles sticking to each other, we measure this height as a function of sample age with the results shown in Fig.~\ref{fig:grtime} for the three samples that form colloidal gels.  In general, the peak heights increase with age, consistent with coarsening (more particles stuck together) and qualitatively matching the observation that the dynamics slow down.  In each experiment the larger particles exhibit more change over time, shown by the large circle symbols in each panel.  The relatively smaller change for the small particles suggests that they are less involved in the restructuring that occurs; or at least, the restructuring does not involve the small particles rearranging to contact other small particles nearly as much as the large particles rearranging to contact other large particles.

\section{Comparison of 2D and 3D data sets} 

The microgravity data is comprised of 2D images of 3D samples.  To better understand the relationship between 2D images and 3D reality, we conduct ground-based experiments as described in Sec.~\ref{sec:groundmethods}.  The dynamics should be isotropic, so we do not expect the mean square displacement measurements to depend on whether the observation is 2D or 3D (which matches prior observations \cite{weeks02sub}).  In contrast, the pair correlation function can differ between 2D and 3D, so we use our data to examine the differences.  

\begin{figure}
\centering
\includegraphics[width=0.9\columnwidth]{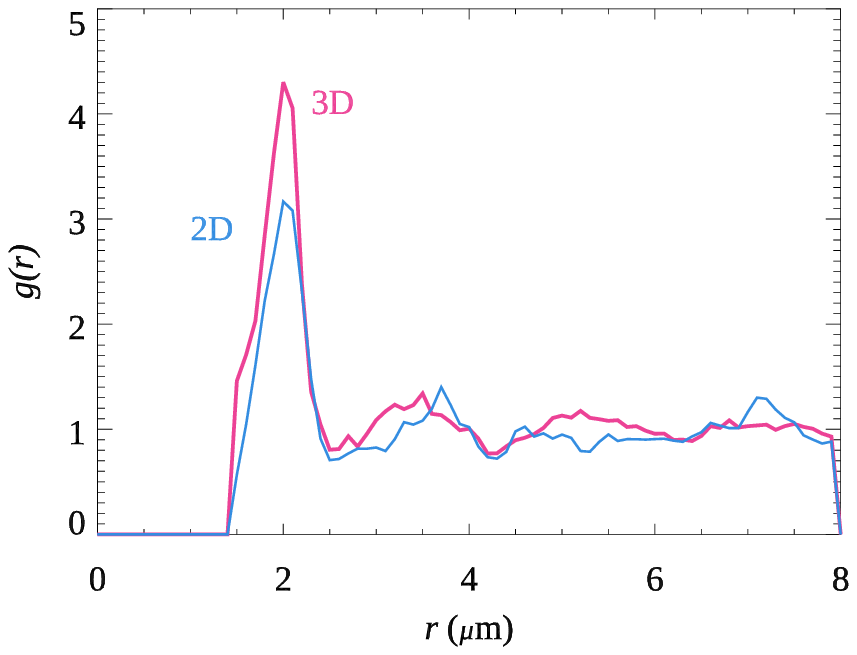}
\caption{\label{gofrground}
The pair correlation function $g(r)$ for a gel sample with a volume fraction $\phi = 0.10$ and a polymer concentration of $c=16$~mg/ml. The thin blue line shows the 2D data and the thicker red line is the 3D data as indicated.
}
\end{figure} 

Accordingly, Fig.~\ref{gofrground} shows an example of $g(r)$ for a gel from the ground experiment, as computed from 2D and 3D analysis.  The gel is the one shown in Fig.~\ref{gel} and has a volume fraction of $\phi = 0.11$ and a polymer concentration of $c=8.5$~mg/ml.  As described in Sec.~\ref{sec:groundmethods}, the 2D and 3D analysis use the same images; the difference is that the 2D analysis treats each image as if it is independent, without considering the context of preceding or subsequent images. Both $g(r)$ curves show a peak at $r \approx 2$~$\mu$m, comparable to the diameter of the particles which is $2a=2.16 \pm 0.02$~$\mu$m.  This is reasonable as the sample is a gel, and the most likely particle separation distance should be $r \approx 2a$ indicating that the particles are in contact. Note that $g(r)>0$ for $r<2a$ reflects particle identification errors; in particular, two particles in contact have slightly overlapping images and this moves their apparent centers-of-mass closer together.

\begin{figure}
\centering
\includegraphics[width=0.9\columnwidth]{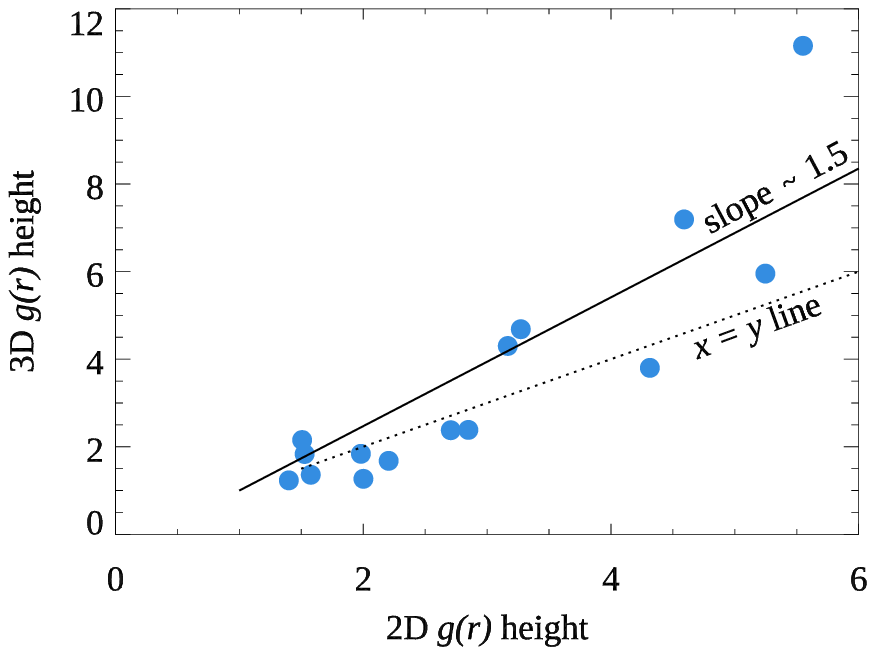}
\caption{\label{maxgr}
Graph of the peak height of $g(r)$ in 2D and 3D from all the samples in the ground experiments.  The dotted line indicates $x=y$; the data above this line indicate that the 2D $g(r)$ height is smaller than the real value calculated from the 3D data.  The solid line is a best fit constrained to go through the point (1,1), with a slope equal to 1.47.  The data points have an uncertainty of about $\pm 40$\% due to choices made in the particle identification, and choices made in calculating $g(r)$ (in particular, the width of the binning in $r$). 
}
\end{figure}

To evaluate the pair correlation function across all ground-based samples, we use the peak height of $g(r)$ to establish a comparison between 2D and 3D. Figure~\ref{maxgr} shows the peak height of $g(r)$ for each sample.  The plot reveals a sensible correlation: the higher the peak in 2D, the higher it is in 3D. The noise in the data reflects the spatially heterogeneous nature of the samples. In general, the peak height of $g(r)$ in 3D is typically greater than that in 2D. Although we believe the trend is true, there are several considerations that affect the coefficient of the linear correlation presented in Fig.~\ref{maxgr}. 

\begin{figure}
\centering
\includegraphics[width=1\columnwidth,trim=0.5in 0.5in 0.8in 0.0in]{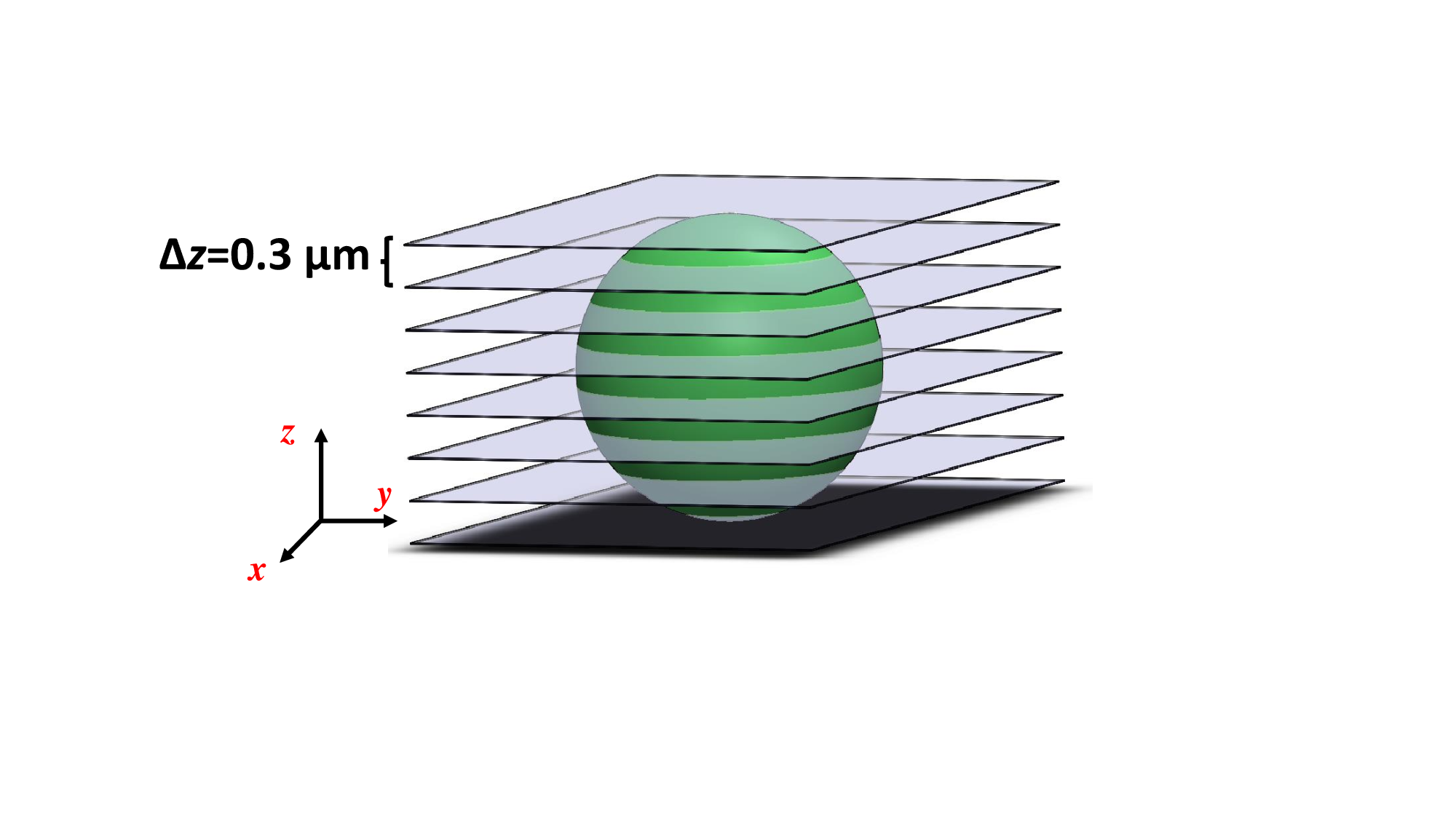}
\caption{\label{sphere}
A schematic showing the relationship between 2D and 3D imaging for one particle. Each layer represents a 2D image, taken a distance 0.3 microns apart. As the focal plane varies, a $z$ stack of these slices creates the 3D image. In this scenario, the same particle is identified in several slices. 
}
\end{figure}

\begin{figure*}
\centering
\includegraphics[width=1.9\columnwidth,trim=0.5in 3in 1.5in 0.0in]{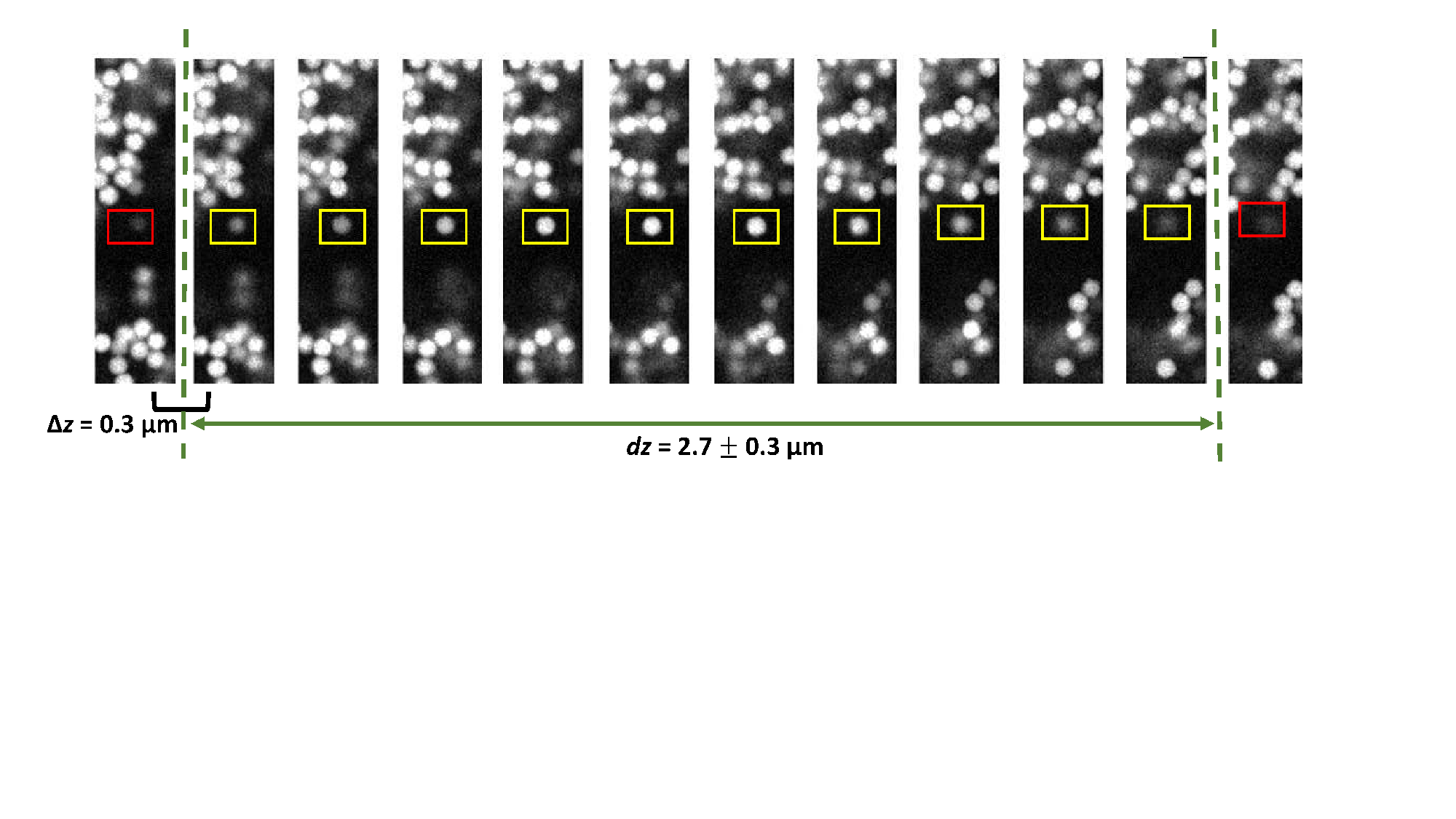}
\caption{\label{dz}
This image sequence is taken from a 3D image stack, scanning through the sample in the $z$ direction.  From left to right, the images are taken deeper in the sample.  The yellow frames represents the z-slices in which the highlighted particle is identified and counted.  The first and last images where the particle is highlighted in red represent that the particle doesn't meet identification criteria. The distance between slices is $\Delta z=0.3$ microns and the effective range is $dz = 2.7 \pm 0.3$~$\mu$ over which the particle is ``visible'' to the software.
}
\end{figure*}

First, we believe that there is an overestimation of the peak $g(r)$ in 2D that is attributed to a higher apparent area fraction compared to the true volume fraction in 3D. This discrepancy arises because particles have a finite diameter (in this case 2.16~$\mu$m) so they overlap the focus of the microscope if their centers are within $\sim 1$~$\mu$m, see the sketch in Fig.~\ref{sphere}.  Thus, particles appear in multiple 2D slices within the 3D image stack \cite{dehoog01}.  This is highlighted in Fig.~\ref{dz}, which shows a sequence of successive $z$-slices from a 3D image stack.  The highlighted particle is identified as a particle using 2D image analysis in all but the first and last image of this sequence.  The fact that there are multiple slices of the same particle is helpful for 3D particle tracking \cite{dinsmore01}, but when analyzing these images in 2D, it means that particles all appear multiple times and thus increase the apparent area fraction.

Additionally, the particle tracking software ensures that particles are not closer than a specified tolerance, typically slightly less than one particle diameter.  As a result, the particles identified from 2D images appear to have a higher area fraction. This effect is similar to compressing a thick slice of the 3D gel ($\approx 2.7$~$\mu$m thick, as suggested by Fig.~\ref{dz}) into a 2D image.  The apparent higher area fraction therefore increases the peak height of $g(r)$.  In contrast, the full 3D image accurately identifies each particle as each one appears only once in the 3D space. 
In reality, the precise identification (or not) of a particle's position in a 2D image is influenced by several factors. These include optical characteristics such as the microscope objective's depth of focus, the Numerical Aperture of the objective lens, and the success of index of refraction matching between particles and solvent. Other factors depend on the particle tracking parameters used.  For example, in Fig.~\ref{dz}, one could set a high brightness threshold for particle identification, limiting the identified particles to only central images. In practice, we select the parameters that best identify particles that match our judgment of what is seen in the images \cite{crocker96}, which applies to our analysis of both ground-based images and ACE-M1 microgravity images. Due to variations in optics and sample conditions, direct comparison between ground-based and space-based images is not feasible.  Despite these challenges, the linear relationship between 2D and 3D peak heights suggests a consistent correlation. This specific ratio will be different and unknowable for the microgravity images, but the fact that the trend appears fairly robust across these samples is encouraging. It suggests that the relative features of the microgravity data seen in Figs.~\ref{gofrnasa} and \ref{fig:grtime} are correct.

\section{Conclusions}

The analysis of NASA's ACE-M-1 colloidal gels provides valuable insights into the aging processes of these materials under microgravity conditions. The observed slowdown in dynamics aligns with the gel coarsening process, where gel strands thicken, and the number of mobile particles decreases. 

Despite the low particle size ratio of $a_L/a_S \approx 1.2$, we observe distinctly different behaviors between large and small particles. Although the diffusivity is expected to scale with size, with larger particles diffusing more slowly than smaller ones, the situation differs for depletion forces. The larger particles experience a stronger depletion force, leading to a larger attractive force between two large particles in contact, $(a_L/a_S)^2 \approx 1.49$ times stronger than between two small particles \cite{asakura54,kaplan94,dinsmore96,yodh01}. This small size difference results in significant behavioral variations, with large particles more firmly incorporated into the gel even at early times and their $g(r)$ showing a higher first-neighbor peak, indicating a prevalence of large-large contacts.

Moreover, we confirm that stronger attractive forces accelerate the coarsening process, significantly impacting the stability and time evolution of colloidal gel structures. Our observations indicate that the temporal evolution of these gels arises from two key factors: the incorporation of smaller, more mobile particles into the gel and the gradual restructuring of the gel, leading to an increase in the number of nearest neighbors. This restructuring is evidenced by the growing height of the first peak in $g(r)$ with time for the gel samples. These structural changes are associated with a slow power-law decay in the mean motion of particles within the gel, observed over nearly two orders of magnitude of time scales in these microgravity experiments.

The comparison between 2D and 3D datasets highlights the complexities of interpreting 2D microgravity images, particularly with respect to apparent differences in the pair correlation function $g(r)$ results. The microgravity movies consist of 2D projections of 3D samples, and our ground-based data show a correlation between the peak heights of $g(r)$ in 2D and 3D observations. Although we cannot definitively determine whether the true 3D $g(r)$ structures are more pronounced than those observed in the microgravity data, we can confirm that the systematic changes in $g(r)$ seen in the microgravity images are real.

We thank J.~Crocker, M.~Lynch, and T.~Kodger for helpful discussions, and W.~Poon and A.~Schofield for providing the colloidal particles for the ground-based experiments.  This work was supported by NASA (80NSSC22K0292), and by the Emory University Emory Integrated Cellular Imaging Core Facility (RRID:SCR{\_}023534).

\end{document}